\documentclass[aps,pra,twocolumn]{revtex4-1}

\usepackage{amsmath}
\usepackage{amssymb}
\usepackage{mathtools}
\usepackage{bbm}

\usepackage{setspace}
\usepackage{bm}
\usepackage{graphicx}
\numberwithin{equation}{section}
\usepackage{hyperref}

\begin{document}


\title{Hamiltonian Tomography of Photonic Lattices}
\author{Ruichao Ma, Clai Owens, Aman LaChapelle, David I. Schuster \& Jonathan Simon}
\affiliation{James Franck Institute and Department of Physics at University of Chicago}
\date{\today}

\begin{abstract}
In this letter we introduce a novel approach to Hamiltonian tomography of non-interacting tight-binding photonic lattices. To begin with, we prove that the matrix element of the low-energy effective Hamiltonian between sites $i$ and $j$ may be obtained directly from $S_{ij}(\omega)$, the (suitably normalized) two-port measurement between sites $i$ and $j$ at frequency $\omega$. This general result enables complete characterization of both on-site energies and tunneling matrix elements in arbitrary lattice networks by spectroscopy, and suggests that coupling between lattice sites is actually a \emph{topological} property of the two-port spectrum. We further provide extensions of this technique for measurement of band-projectors in finite, disordered systems with good flatness ratios, and apply the tool to direct real-space measurement of the Chern number. Our approach demonstrates the extraordinary potential of microwave quantum circuits for exploration of exotic synthetic materials, providing a clear path to characterization and control of single-particle properties of Jaynes-Cummings-Hubbard lattices. More broadly, we provide a robust, unified method of spectroscopic characterization of linear networks from photonic crystals to microwave lattices and everything in-between.
\end{abstract}

\maketitle

\section{\label{sec:Intro} Introduction}
The curse and blessing of synthetic quantum materials is the control these systems afford. This control enables access to near-arbitrary lattice geometries \cite{jo2012,tarruell2012,Greiner2002,gomes2012}, tunable interaction range \cite{gonzalez2015}, and all variety of state/phase preparation and readout techniques \cite{Bakr2009,gericke2008,husmann2015}. The challenge is that every added degree of control provides another opportunity for disorder to creep in, substantially altering the anticipated manybody physics. A variety of approaches have been developed to control disorder, ranging from projection of corrective potentials onto cold atoms \cite{Bakr2010} to improving lattice fabrication in superconducting circuits \cite{underwood2012} and 2DEGs \cite{Stormer1999}. Indeed, as fabrication techniques have improved in 2DEGs, the accessible fractional hall landscape has opened for study of immense array of exciting topological phases, and it seems other synthetic material systems could follow a similar trend.

If disorder is to be corrected site-by-site, it must be characterized locally. This task is challenging, because information about the onsite energy of a lattice site and its tunneling rates to its neighbors are encoded non-trivially (and apparently non-locally) in the eigen-value/vector spectrum of the system. In the case of a 1D tight-binding chain, the reflection spectrum off of the system end is sufficient to extract the full non-interacting Hamiltonian (see \cite{1DTomography2009} and Appendix \ref{app:1DChain}). For a 2D lattice of known topology, it is possible to make measurements along a 1D boundary to extract the Hamiltonian parameters \cite{Maruyama2009}, with sufficiently high signal-to-noise. Here we point out a unique opportunity to employ direct spectroscopic tools to extract particular desired matrix-elements of the single-particle Hamiltonian. We describe a general technique for resolving matrix elements of an arbitrarily connected Hamiltonian between lattice sites via simple two-port transmission and one-port reflection (local density of states) measurements; we then extend the technique to measurement of band projectors and Chern numbers.

\section{\label{sec:Theory} Theory of Lattice Spectroscopy}
\subsection{Formulae for Arbitrary Linear Networks}
Suppose that we would like to characterize a non-interacting network of lattice-sites in the site-basis, by answering specific questions like ``what is the energy cost to put a particle on site $i$?'' or ``what is the tunnel-coupling between sites $i$ and $j$?''. One might attempt to characterize the \emph{full lattice} by performing two-port measurements between all pairs of sites $(m,n)$, and then fitting the results with an analytic model to extract the underlying lattice parameters. This works in principle, but generally is highly susceptible to noise and requires $O(N^2)$ measurements (except in the 1D case, see Appendix \ref{app:1DChain}); here we prove that the information for matrix elements of the Hamiltonian $H_{ij}$ is entirely encoded in the frequency-dependent two-port measurement $S_{ij}(\omega)$ between only the two sites $i$ and $j$ of interest.

Let the system Hamiltonian be given by (in what follows we set $\hbar=1$): 
\begin{equation}
H=\sum_{l}(\omega_l+i\kappa_l/2) a_l^\dagger a_l-\sum_{i\neq j}t_{ij}a_i^\dagger a_j 
\end{equation}
Where $t_{ij}$ is the direct tunnel-coupling between sites $i$ and $j$, $\omega_l$ is the energy cost to place a photon on site $l$, and $\kappa_l$ is the lifetime of a particle on site $l$. We have employed a non-Hermitian Hamiltonian formalism which applies in the weak-driving limit \cite{cohen1992,sommer2015quantum}.  It is straightforward to show that in this weak driving limit, the resonator transmission between sites $\alpha$ and $\beta$ at frequency $\omega$ is given by:
\begin{equation}
S_{\alpha\beta}(\omega)=\sqrt{\kappa_c^\alpha \kappa_c^\beta} \times \langle \alpha | \frac{1}{\omega-H_c^{\alpha\beta}} | \beta \rangle
\end{equation}

Here $H_c^{\alpha\beta}\equiv H+i\frac{\kappa_c^\alpha}{2} a_{\alpha}^\dagger a_{\alpha}+i\frac{\kappa_c^\beta}{2} a_{\beta}^\dagger a_{\beta}$ is the Hamiltonian adjusted for loss due to in-/out- coupling employed for probing the photonic network at sites $\alpha$ ($\kappa_c^\alpha$) and $\beta$ ($\kappa_c^\beta$), and $|i\rangle$ is the quantum state with a single photon at site $i$.

Consider $\mathit{Pr}(\int \omega S_{ij}(\omega)\mathrm{d}\omega)$, which diverges without the Cauchy principal value $\mathit{Pr}()$. To perform the integration, we employ the following definitions: $|\mu\rangle$ is the single-photon eigenstate of $H_c^{\alpha\beta}$ with eigenvalue $\epsilon_\mu$, and $\langle\nu|$ is the element of the dual space to $|\mu\rangle$ defined such that $\langle\nu|\mu\rangle=\delta_{\mu\nu}$; note that $\langle\nu|\neq[|\nu\rangle]^\dagger$, because $H_c^{\alpha\beta}$ is not Hermitian, so the matrix of eigenvectors is not unitary. We can then write:
\begin{align}
&\int\omega S_{\alpha\beta}(\omega)\mathrm{d}\omega\nonumber\\
=&\int\omega\sqrt{\kappa_c^\alpha \kappa_c^\beta} \times \langle \alpha | \frac{1}{\omega-H_c^{\alpha\beta}} | \beta \rangle\,\mathrm{d}\omega \nonumber\\
=&\sqrt{\kappa_c^\alpha \kappa_c^\beta}\int\omega\langle \alpha | \sum_\mu \frac{|\mu\rangle\langle\mu|}{\omega-\epsilon_\mu} | \beta \rangle\,\mathrm{d}\omega \nonumber\\
=&\sqrt{\kappa_c^\alpha \kappa_c^\beta}\langle \alpha | \sum_\mu \left[\int\omega \frac{|\mu\rangle\langle\mu|}{\omega-\epsilon_\mu} \mathrm{d}\omega\right]|\beta\rangle \nonumber\\
=&\sqrt{\kappa_c^\alpha \kappa_c^\beta}\langle \alpha | \sum_\mu \left[\int(1+\frac{\epsilon_\mu}{\omega-\epsilon_\mu})|\mu\rangle\langle\mu| \mathrm{d}\omega\right]|\beta\rangle \nonumber\\
=&\sqrt{\kappa_c^\alpha \kappa_c^\beta}\langle \alpha | \left[W+\sum_\mu |\mu\rangle\epsilon_\mu\langle\mu|\right]|\beta\rangle \nonumber\\
=&\sqrt{\kappa_c^\alpha \kappa_c^\beta}(W\langle \alpha |\beta\rangle+i\pi\langle \alpha | H_c^{\alpha\beta}|\beta\rangle)
\end{align}

Here $W$ is the range of integration. To extract the coupling strengths $\kappa_c^{\alpha,\beta}$, we must also measure the 1-port reflections $S_{\alpha\alpha}(\omega)$ and $S_{\beta\beta}(\omega)$. A simple calculation reveals that $\int S_{\alpha\alpha}(\omega)\mathrm{d}\omega=i \pi \kappa_c^\alpha$, thus we may finally write:
\begin{equation}
\label{eqn:termextract}
\langle \alpha | H_c^{\alpha\beta}|\beta\rangle=\frac{\int\omega S_{\alpha\beta}(\omega)\mathrm{d}\omega}{\sqrt{\left(\int S_{\alpha\alpha}(\omega)\mathrm{d}\omega\right)\left(\int S_{\beta\beta}(\omega)\mathrm{d}\omega\right)}}-\frac{W\langle \alpha |\beta\rangle}{i\pi}
\end{equation}

Thus we see that the matrix element of the Hamiltonian that couples a single photon in site $\alpha$ to site $\beta$ is given by the expectation of frequency weighted by the two-port measurement (as measured by a vector network analyzer, for example) between those two sites, properly normalized by one-port reflection measurements. If $\alpha\neq\beta$, then such a measurement provides the tunneling matrix element $t_{\alpha\beta}$, including its phase. If $\alpha=\beta$, this is an offset-subtracted reflection measurement, and it results in $\omega_\alpha+i(\kappa_\alpha+\kappa_c^\alpha)/2$, the onsite energy at site $\alpha$, with the imaginary part providing the coupled resonator linewidth. For sites which are not directly connected, the measurement will result in a zero value. It is somewhat surprising that sites which are coupled through the network, though not directly, yield zero for the integral -- this suggests that there is a hidden topological property in the frequency-dependent two-point measurement between non-directly-connected sites.

\begin{figure}
\includegraphics[width=80mm]{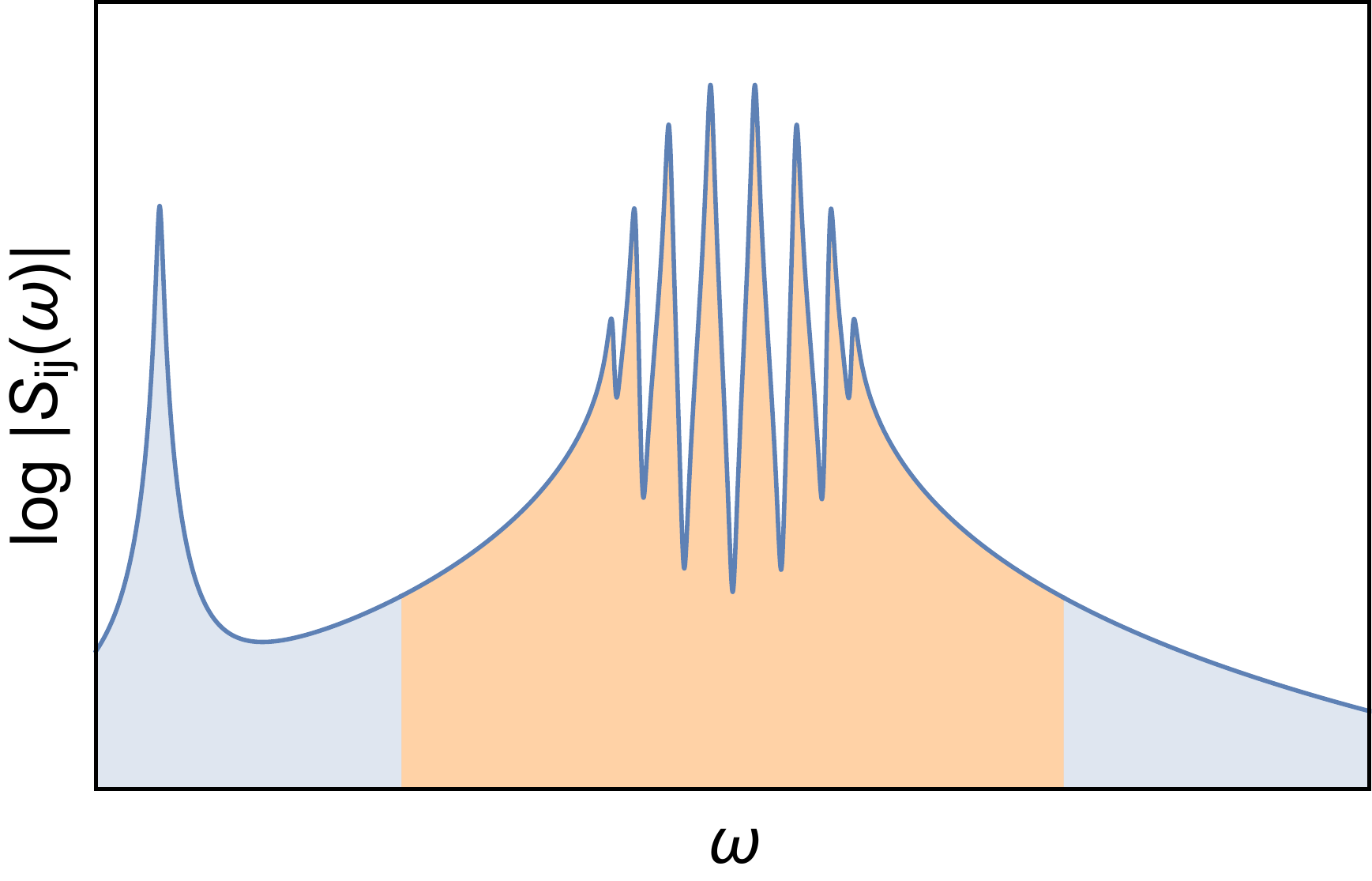}
\caption{\label{Figure:Truncation} \textbf{Truncation of integration region}. To avoid the divergence in the integration of the tails of the transmission and reflection spectra, the symmetric tails of the spectra should be identified by the locations in the spectra where the lattice response has reduced to a simple Lorentzian, decaying as $1/\Delta$ (for $S_{ii}$; the decay is $1/\Delta^2$ for $S_{ij}$, $i\neq j$), and their values are identical. Here $\Delta$ is the detuning of the probe frequency to the manifold of resonances being considered. The integration then should only be performed between these two points (shown in tan), and the divergence of the tails (shaded light blue) will then cancel. In practice, choosing the cutoff location is a trade-off between ensuring that one is far enough from the resonant features to be in regions with $1/\Delta$ (or $1/\Delta^2$) decay, but not so far out that other resonator modes (corresponding to parasitic resonances outside of the effective model as shown at left, or other bands within the effective model) become important. The impact of these other modes on the integration can be further reduced using the technique outlined in Appendix \ref{app:MultiCouple}.}
\end{figure}

The power of this approach is clear: even with a tremendous number of modes (approaching a continuum), the bare frequency of a single resonator, or the tunnel coupling between a pair of resonators, can be directly extracted from 1- or 2- port frequency dependent measurements. This provides a robust linear method for estimating matrix elements of the Hamiltonian that is much less sensitive to noise than other methods involving e.g. fitting of all coupled modes. Handling the logarithmic divergence of the integral (formally taken care of via a Cauchy principal value) requires some care, however, and we suggest two approaches:

\begin{enumerate}
\item In small lattices, where the individual normal modes are spectrally resolved, the integrals may be performed by identifying and fitting the individual resonances in the one/two port measurements, and then evaluating the integrals as sums over said resonances (here $A_l^{\alpha\beta}$, $\phi_l^{\alpha\beta}$, $\omega_l^{\alpha\beta}$ $\gamma_l^{\alpha\beta}$ are the parameters resulting from the fit to the observed two-point spectrum between sites $\alpha$ and $\beta$, $S_{\alpha\beta}^{\mathit{obs}}(\omega)$):
\begin{alignat}{2}
\hspace{7mm}&S_{\alpha\beta}^{\mathit{obs}}(\omega) &&=\sum_l \frac{A_l^{\alpha\beta} e^{i \phi_l^{\alpha\beta}}}{1+i \frac{(\omega-\omega_l^{\alpha\beta})}{\gamma_l^{\alpha\beta}/2}} \nonumber\\
&\textbf{N}^{\alpha} &&\equiv \int S_{\alpha\alpha}^{\mathit{obs}}(\omega)\mathrm{d}\omega\nonumber \\
& &&=\frac{\pi}{2} \sum_l A_l^{\alpha\alpha} e^{i \phi_l^{\alpha\alpha}}\gamma_l^{\alpha\alpha}\nonumber \\
&\textbf{X}^{\alpha\beta} &&\equiv \int \omega S_{\alpha\beta}^{\mathit{obs}}(\omega)\mathrm{d}\omega\nonumber-\frac{W\langle \alpha |\beta\rangle}{i\pi}\sqrt{\textbf{N}^{\alpha}\textbf{N}^{\beta}} \nonumber\\
& &&=\frac{\pi}{2} \sum_l A_l^{\alpha\beta} e^{i \phi_l^{\alpha\beta}}\gamma_l^{\alpha\beta}(\omega_l^{\alpha\beta}+i\gamma_l^{\alpha\beta}/2)\nonumber\\
&\langle \alpha | H_c^{\alpha\beta}|\beta\rangle &&=\frac{\textbf{X}^{\alpha\beta}}{\sqrt{\textbf{N}^{\alpha}\textbf{N}^{\beta}}}
\end{alignat}
\item In larger lattices, where the individual modes cannot be spectrally resolved, the integrals may be explicitly computed from the observed spectra, taking care to symmetrically cut off the tails at low- and high- frequencies, to cancel the logarithmic divergence of the integration (see Fig.~\ref{Figure:Truncation}). Note that this cutoff need not be perfect, especially for the normalization terms (coming from reflection measurements), where the divergence is logarithmic. On the other hand, the integration in equation \ref{eqn:termextract} diverges linearly for on-site matrix elements ($\alpha=\beta$) of the Hamiltonian, so it is crucial to subtract off the integration-range dependent correction given by the second term.
\end{enumerate}

Note also that low-area peaks contribute very little to the value of measured Hamiltonian matrix element, so finite signal to noise ratio is likely not a fundamentally limiting factor in the same way that it would be if one attempted to use many transmission and reflection measurements to fully \emph{invert} and extract the lattice Hamiltonian.

\begin{figure}
\includegraphics[width=85mm]{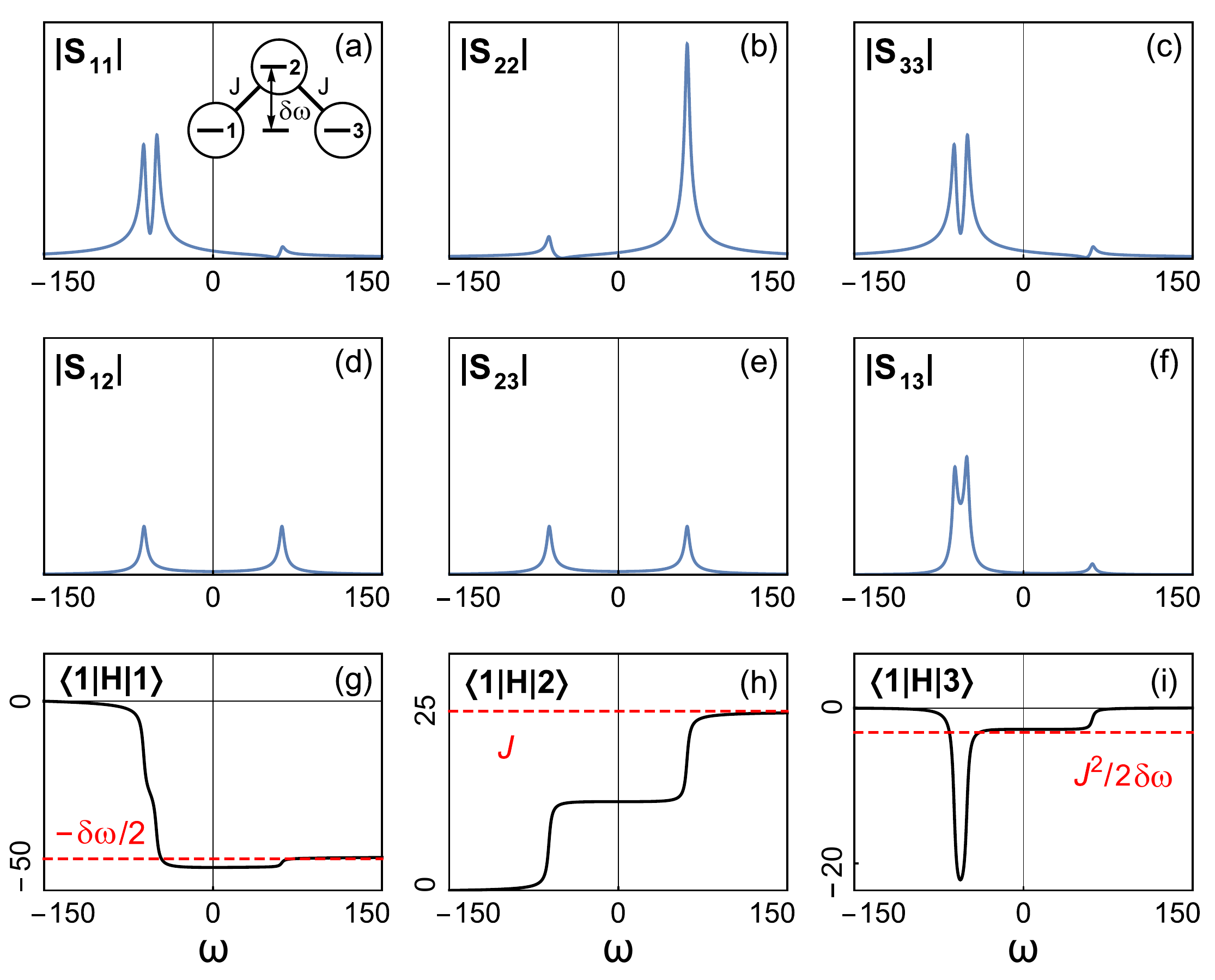}
\caption{\label{Figure:SpecDemo} \textbf{Three site spectroscopy}. (a, inset) shows a three-site tight-binding chain, whose outer sites are at equal energies, while the central site is detuned. (a)-(c) show reflection spectroscopy of sites 1,2, and 3, respectively. (d)-(f) show transmission spectroscopy for 1-2, 2-3, and 1-3, respectively. All spectra are plotted as the absolute value of the amplitude, and share a common (arbitrary) normalization. The spectral features include a low energy doublet resulting from second-order coupling between sites 1 and 3, and a single isolated high-energy feature resulting from the central site. (g)-(i) explore the sensitivity of the Hamiltonian estimation technique on the integration range of the numerator of Equation \ref{eqn:termextract}. We fix the lower limit of integration at -150\,MHz and vary the upper limit between -150\,MHz and 150\,MHz. (g) measures the onsite energy of site 1, which converges to within $J^2/{\delta\omega}$ of $-\,\delta\omega/2$ once the integration range includes the doublet, and the rest of the way once the high-energy feature is included. (h) measures the tunneling matrix element between sites 1 and 2, converging only once all spectral features are included. (i) measures the tunneling matrix element between sites 1 and 3, which is zero in our model. When only the low-energy doublet is included, the result is of order $J^2/{\delta\omega}$, up to corrections from the renormalization of the onsite Wannier function. To obtain the correct value, all spectral features must be included in the integration. For (g)-(i) we plot only the real-part of the estimated elements, as the imaginary part converges less rapidly.}
\end{figure}

As a simple demonstration of this technique, we consider a three-site tight-binding model, as shown in Fig.~\ref{Figure:SpecDemo}(a, inset), where the outer two sites are tuned to frequency $\omega_0-\delta\omega/2$, and the central cite is tuned to frequency $\omega_0+\delta\omega/2$; the outer two sites are coupled to the central cite with a tunneling energy $J$. Figure \ref{Figure:SpecDemo}(a)-(f) show computed reflection and transmission spectra $S_{11}$, $S_{22}$, $S_{33}$, $S_{12}$, $S_{23}$, and $S_{13}$, respectively. As expected the outer two sites hybridize through an effective second-neighbor coupling ${\sim}\,J^2/{\delta\omega}$, while the central site's resonance is detuned by ${\sim}\,\delta\omega$. For $\delta\omega=$100\,MHz, $J=$25\,MHz, Fig.~\ref{Figure:SpecDemo}(g) shows the on-site energy $\langle 1|H|1\rangle$ extracted via the tomography technique from the preceding section, as the upper limit of integration is varied. It is apparent that the site-energy converges to within ${\sim}\,J^2/{\delta\omega}$ of the correct value as soon as the integration region includes the low-energy doublet, and is further corrected as the region passes across the isolated (and small) high-energy resonance. To extract $\langle 1|H|2\rangle$, Fig.~\ref{Figure:SpecDemo}(h) shows the tomography result as a function of the upper limit of integration; once all resonances are included, this value converges to $J$, as anticipated. Finally, Fig.~\ref{Figure:SpecDemo}(i) shows the tunneling matrix element $\langle 1|H|3\rangle$ as a function of the upper limit of the integration; for a range that \emph{only} includes the doublet, the tomography procedure yields a result proportional to the second-order tunneling rate of $J^2/{\delta\omega}$ (though this precise value is not obtained: the probed lattice-sites are not the ``Wannier functions'' of the effective theory once the central site has been adiabatically eliminated, and thus there are corrections, see Appendix \ref{app:MultiCouple}); once the high-energy resonance is included, the true tunneling rate of zero is recovered.

\subsection{Band Projectors and Real-Space Measurement of the Chern Number}
An emerging goal in synthetic topological materials is to characterize their topological invariants. While the Hall conductivity is the method of choice in the solid state, transport measurements can be challenging in synthetic systems, particularly those where the ``charge carriers'' are bosons rather than fermions. Furthermore, such systems are typically subject to both disorder effects, and the impact of finite size/boundaries, both of which break the translational invariance necessary for application of the TKNN formula \cite{TKNN1982} for the Chern invariant. In a seminal work \cite{kitaev2006anyons}, Kitaev proved that the Chern number could be computed for a disordered system, so long as the disorder is small enough that the bands remain spectrally isolated from one another. In this case, one may define a projector into band $\mu$ with matrix elements between lattice sites $i$ and $j$:
\begin{equation}
P^\mu_{ij}\equiv\langle i | \left [ \sum_{n\,\in\,\text{Band}\, \mu}|n\rangle\langle n | \right ] | j \rangle
\end{equation}

\begin{figure}
\includegraphics[width=70mm]{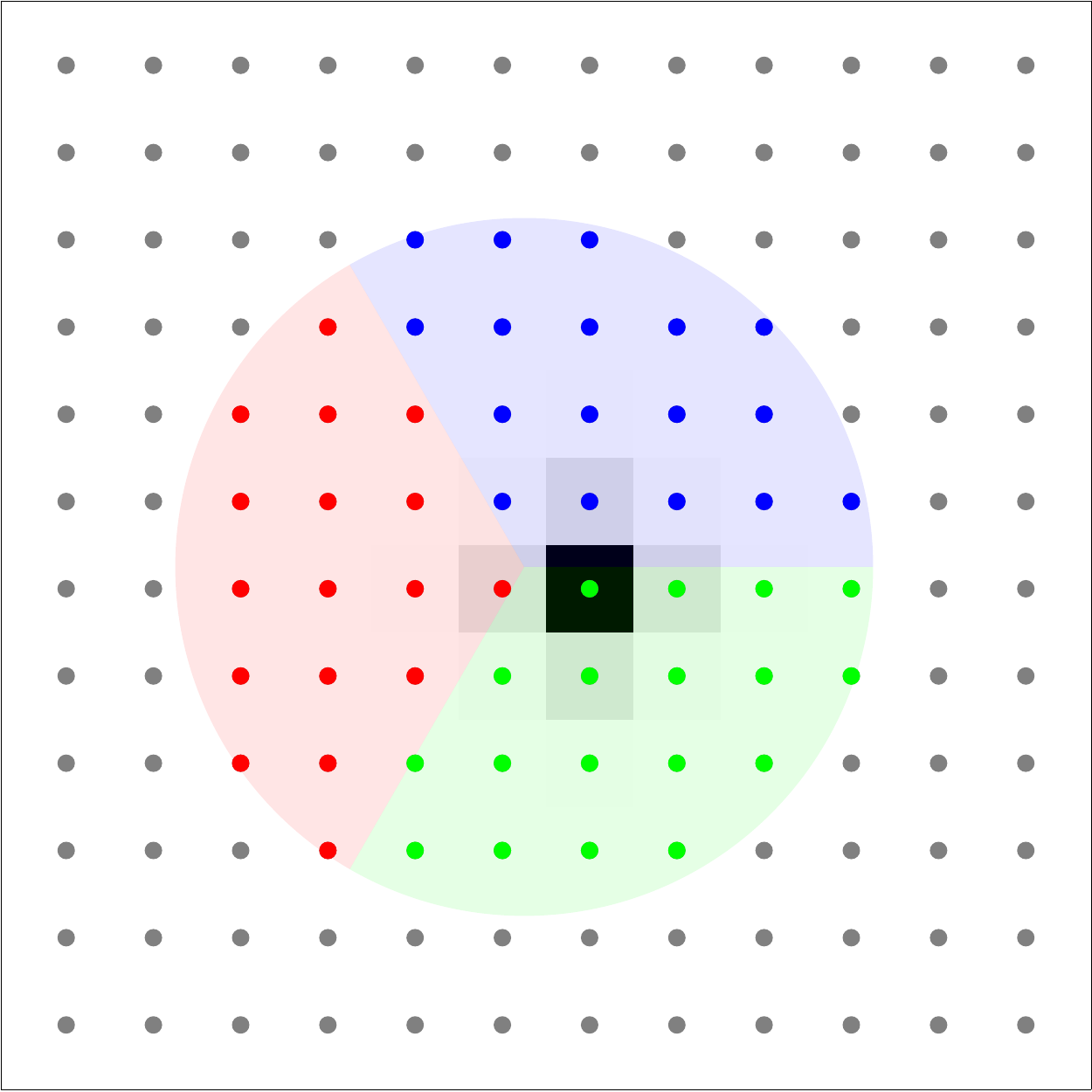}
\caption{\label{Figure:RealSpaceChern} \textbf{Measuring Chern numbers in real space}. To measure the Chern number of a disordered band in the bulk of a Chern insulator, a bulk region large compared to the unit cell size (magnetic length in the Hofstadter model, whose band-projector onto an arbitrary bulk site is shown in gray-scale for $\alpha=\frac{1}{4}$) is partitioned into three similarly sized regions (red, green, and blue). The difference of triple band-projector products $red\rightarrow green\rightarrow blue$ and $blue\rightarrow green\rightarrow red$, summed over all sites in each region, is equal to the Chern number $C/(12\pi i)$. There are a number of ways to spectroscopically measure this projector, discussed in the text.}
\end{figure}

If the sites in the bulk of the system are then partitioned into three non-overlapping but adjacent regions $A,B,C$, as in Fig.~\ref{Figure:RealSpaceChern}, the Chern number may be written:
\begin{equation}
C^\mu=12\pi i \sum_{\mathclap{\alpha\in A, \beta\in B, \gamma\in C}}(P^\mu_{\alpha\beta}P^\mu_{\beta\gamma}P^\mu_{\gamma\alpha}-P^\mu_{\alpha\gamma}P^\mu_{\gamma\beta}P^\mu_{\beta\alpha})
\end{equation}

\begin{figure}
\includegraphics[width=80mm]{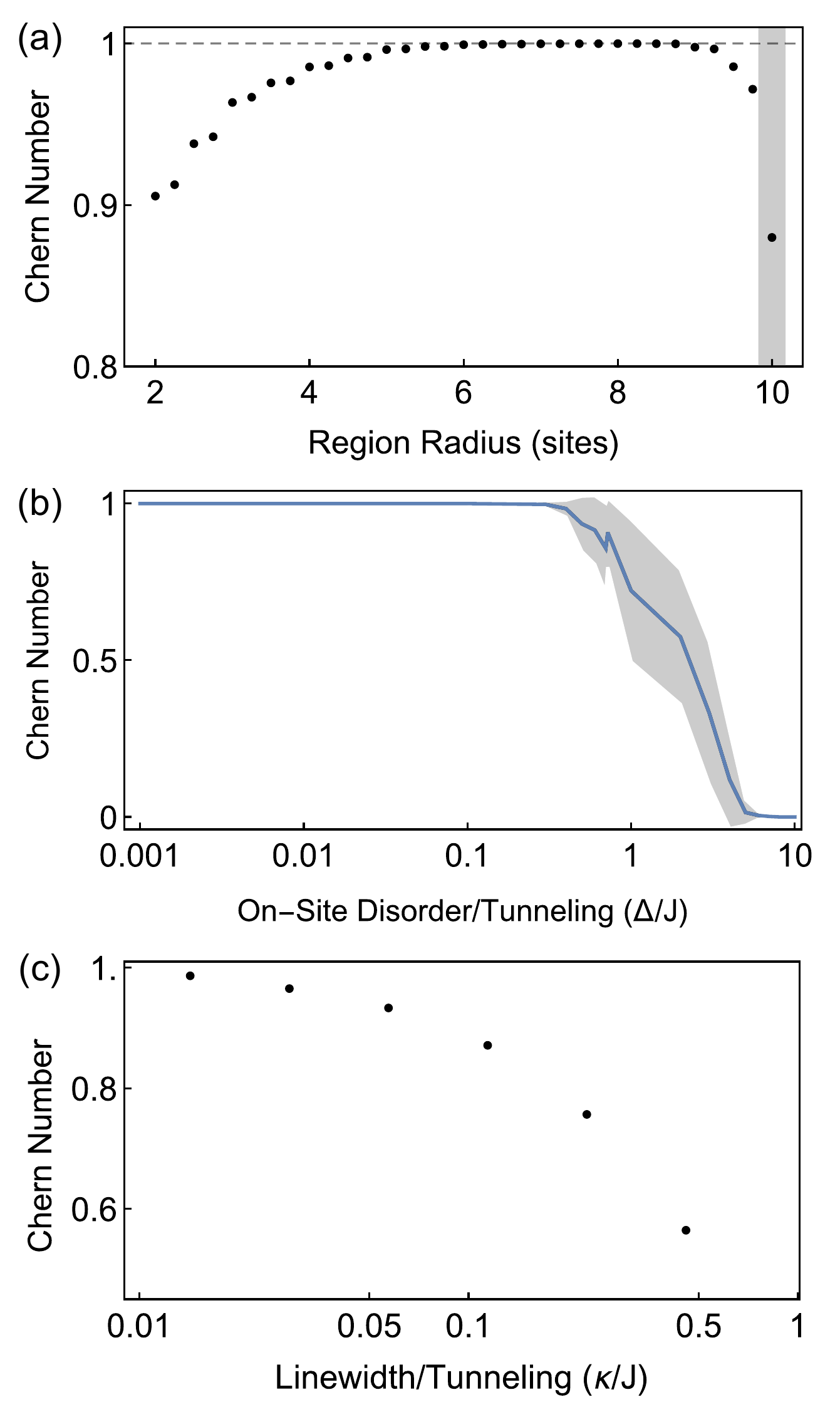}
\caption{\label{Figure:RealSpaceChernMeas} \textbf{Technical limitations of Chern number measurement}. (a) When the Kitaev approach is applied to an array of lossless resonators coupled in an $\alpha=\frac{1}{4}$ Hofstadter configuration, through a numerically computed band-projector, the measurement of the lowest band Chern number depends upon the employed region size. Once the region size becomes larger in radius than the magnetic unit cell ($\frac{1}{\alpha}=4$ sites), the measured value approaches the translationally invariant value $C=1$ from the TKNN expression. It begins to fall off once the region approaches the system edge (gray bar). (b) To explore the sensitivity of the real-space Chern number to disorder, it is plotted versus the site-to-site (random) variation of the on-site energy. The error band shows the variation in the result over realizations of the disorder, and reveals that the Chern number (or at least its real-space estimator) is robust to disorder up to ${\sim}\,0.1\times J$, which is comparable to the band-splitting in the $\frac{\alpha}{4}$ Hofstadter model. (c) The Kitaev approach may be applied to a realistic array of lossy resonators, using a frequency integration of the two port measurement $S_{ij}(\omega)$, as described in the text, for a region with five site radius. In this case, the resonator loss compared with the tunneling rate determines the fidelity. It appears necessary that loss rate $\leq 0.03 \times$tunneling rate to achieve fidelity $\geq 0.95$.}
\end{figure}

While the regions A, B, and C must be infinitely large to ensure precise convergence of the Chern number to the TKNN invariant defined from the band structure, in practice a region which is several unit cells (or equivalently magnetic unit cells, in the case of the Hofstadter model) is sufficient to achieve reasonable convergence (at the ${\sim}\,99\%$ level, see Fig.~\ref{Figure:RealSpaceChernMeas}(a). Furthermore, it is essential that $A,B$ and $C$ avoid the system edges, as these provide a contribution to $C_\mu$ which precisely cancels that of the bulk. This approach may be understood as a a direct measurement of the non-reciprocity of the system, as it compares $A\rightarrow B\rightarrow C$ coupling to $C\rightarrow B\rightarrow A$ coupling, similar to the case in a Faraday isolator \cite{Faraday1953}. As shown in Fig.~\ref{Figure:RealSpaceChernMeas}(b), as long as the disorder is an order of magnitude smaller than the band spacing, Chern number quantization is preserved.

The challenge then is to measure the band-projector using the spectroscopic tools at our disposal. We suggest three approaches:
\begin{enumerate}
\item Consider the integral:
\begin{align*}
M^\mu_{ij} &\equiv \mathit{Pr}[\int_{\omega\,\in\,\text{Band}\, \mu}d\omega S_{ij}(\omega)]\\
&=\mathit{Pr}[\int d\omega \langle i |\frac{1}{\omega-H}|j\rangle]\\
&=\mathit{Pr}[\int d\omega\sum_n\frac{\langle i | n\rangle\langle n | j \rangle}{\omega-\epsilon_n}]
\end{align*}
Assuming good band flatness $\frac{\text{Band Width}}{\text{Band Spacing}}\ll 1$ \cite{neupert2011fractional,yao2012topological}, we can integrate across band $\mu$ without accruing a substantial contribution from the other bands, yielding $M^\mu_{ij}\approx i \pi\langle i |\left [ \sum_{n\,\in\,\text{Band}\, \mu}{|n\rangle\langle n |}\right ] | j \rangle$. Therefore $P^\mu_{ij}\approx\frac{1}{i\pi}M^\mu_{ij}$. It is thus sufficient to integrate $S_{ij}(\omega)$ over a single energy-band $\mu$ to extract the matrix element of the projector onto band $\mu$ between sites $i$ and $j$. This integral is only logarithmically sensitive to the limits of integration, so precise cancellation of the tail contributions from finite linewidth and imperfect flatness are possible at near unity fidelities.
\item Consider localized excitation at site $i$ within the bulk of the lattice, at an energy $\hbar \omega_o$ detuned from band $\mu$ by an amount large compared to its width, but small compared with its detuning to other bands. The response at site $j$ is given by $S_{ij}(\omega_o)=\sum_n\frac{\langle i | n\rangle\langle n | j \rangle}{\omega_o-\epsilon_n}$. In the limit that the detuning to all other bands is large, their contribution may be discarded. If at the same time the detuning to band $\mu$ is large compared with the bandwidth, all energy denominators are approximately constant $\omega_o-\epsilon_n\approx\Delta$ for $n\in \text{ Band } \mu$. Then we have $S_{ij}(\omega_o)\approx\frac{1}{\Delta}\sum_{n\,\in\,\text{Band}\, \mu}\langle i | n\rangle\langle n | j \rangle$, and thus $P^\mu_{ij}\approx \Delta \times S_{ij}(\omega_o)$.
\item Consider a localized excitation at site $i$ within the bulk of the lattice, with a temporally short wave-packet energetically centered on band $\mu$. If this pulse can be made short compared to the width of band $\mu$, while simultaneously long enough to not excite other bands, the response of the system immediately after the pulse will reflect the projector onto site $i$ in band $\mu$; if the pulse is insufficiently short compared with the bandwidth of band $\mu$, the excitation will evolve spatially before the pulse has terminated, and the projector cannot be extracted.
\end{enumerate}

The second and third approaches impose \emph{much} more stringent requirements on the band flatness than the first, and as such will not work well for Hofstadter models at high flux per plaquette.

In any of these approaches, it should be possible, in the low-disorder limit, to make use of the approximate translational invariance from one magnetic unit cell to the next to reduce the number of measurements from ${\sim}\,N^2$, where $N$ is the number of sites in one of the regions $A,B,C$, to ${\sim}\,q\times N$, where $q$ is the number of sites within the magnetic unit cell (equal to 4 for $\alpha=\frac{1}{4}$). Because $N\sim\,q^2$, the total number of two-point spectra required to extract the Chern number is thus ${\sim}\,q^3$.

A more fundamental limit comes from the finite lifetime of a photon in the lattice, which provides an additional form of (dissipative) time-reversal symmetry breaking that competes with the topology of the lattice, making $C_\mu$ complex (for precisely flat bands, it may be possible to precisely cancel this contribution through matching the logarithmically diverging tails of the two-point integrals). As shown in Fig.~\ref{Figure:RealSpaceChernMeas}(c), the Chern number can be measured with fidelity above $95\%$ so long as the tunneling rate is $30\times$ the photon decay rate, for an $\alpha=\frac{1}{4}$ Hofstadter model, in spite of the substantial band curvature. The requirement on tunneling compared to decay is consistent with the particle needing time to explore an area whose radius is the magnetic length ${\sim}\,q$, to be sensitive to the Chern number.

\section{\label{sec:Outlook} Outlook}
We have provided a novel toolset for characterizing photonic lattices using one- and two- point measurements to resolve elements of the Hamiltonian. We have further introduced a recipe to extract the band projector, allowing direct measurement of Chern number in real-space. While the proposed approach is designed for photonic lattices where network analyzer technology is commercially available, it can be applied much more broadly to explore properties of coupled quantum dots, acoustical systems, and potentially even electronic systems by reinterpreting STM measurements.

\section{\label{sec:Acknowledgemends}Acknowledgements}
We would like to thank Brandon Anderson, William Irvine, Charles Kane, Michael Levin, Nathan Schine, and Norman Yao for fruitful discussions. This work was supported by ARO grant W911NF-15-1-0397. D.S. acknowledges
support from the David and Lucile Packard Foundation; R.M.
acknowledges support from the University of Chicago MRSEC program of the NSF under grant NSF-DMR-MRSEC 1420709; C.O. is supported by the NSF GRFP.

\appendix
\section{\label{app:MultiCouple}Coupling to Multiple Sites}
In practice, one must be careful to avoid accidental \emph{direct} coupling to multiple lattice sites when performing the spectroscopy of a tunnel-coupled lattice system. Such direct couplings arise naturally because in any real lattice the Wannier functions are not perfectly localized to individual lattice sites. This non-local tail \emph{means} that if the in- and out- couplers are physically connected only to individual sites, they will drive and measure multiple lattice sites.

To understand the consequences of this, consider two degenerate sites at energy $\hbar \omega_0$, $|a\rangle$ and $|b\rangle$, that are tunnel-coupled with an energy $\hbar J$, such that the Hamiltonian in the 1-excitation manifold is $H_0/\hbar=\omega_0(|a\rangle\langle a|+|b\rangle\langle b|)-J(|a\rangle\langle b|+|b\rangle\langle a|)$. Now we drive with a coupler $|\mu\rangle\equiv\cos\epsilon|a\rangle+\sin\epsilon|b\rangle$ (predominantly connected to site a), and measure with coupler $|\nu\rangle\equiv\cos\epsilon|b\rangle+\sin\epsilon|a\rangle$ (predominantly connected to site b), corresponding to a Wannier overlap of ${\sim}\,\epsilon^2$ on adjacent sites.

We then measure $S_{\mu\mu}(\omega)$, $S_{\nu\nu}(\omega)$, and $S_{\mu\nu}(\omega)$, and attempt to extract the Hamiltonian matrix elements. Applying the spectroscopy techniques from the text yields: $\langle\mu|H_0|\mu\rangle_{Spec}=\langle\nu|H_0|\nu\rangle_{Spec}=\hbar\left[\omega_0-J\sin2\epsilon\right]$ and $\langle\mu|H_0|\nu\rangle_{Spec}=-J+(\omega_0-i W/\pi)\sin2\epsilon$. We anticipated that $S_{\mu\mu}(\omega)$ and $S_{\nu\nu}(\omega)$ would provide on-site energies, while $S_{\mu\nu}(\omega)$ was to provide the tunneling energy. In reality, we find that the on-site energy experiences a small correction from the tunneling energy, which, in the tight-binding limit (where $\epsilon\ll1$), is almost certainly negligible. By contrast, the error in the tunneling energy may be \emph{much} larger than $J$ itself if $\epsilon\geq\frac{J}{\omega_0}$.

To circumvent this systematic issue, the measurements of $\langle\alpha|H|\beta\rangle$ may be re-orthogonalized using a basis transformation based upon the matrix $\int\mathrm{d}\omega S_{\alpha\beta}(\omega)$. A simpler solution is to shift all frequencies by some constant $\Omega\,{\sim}\,\omega_0$, and then employ $\tilde{S}_{\mu\nu}(\omega)=S_{\mu\nu}(\omega-\Omega)$ for all resolvent calculations. We are then measuring matrix elements of $H_0-\mathbbm{1}\Omega$, and thus the error in the measurement of $J$ will be of order $(\omega_0-\Omega)\sin\epsilon\leq J\sin\epsilon$, and thus small.\\

\section{\label{app:1DChain}Special Case of a Finite 1D Chain}
\begin{figure}
\includegraphics[width=85mm]{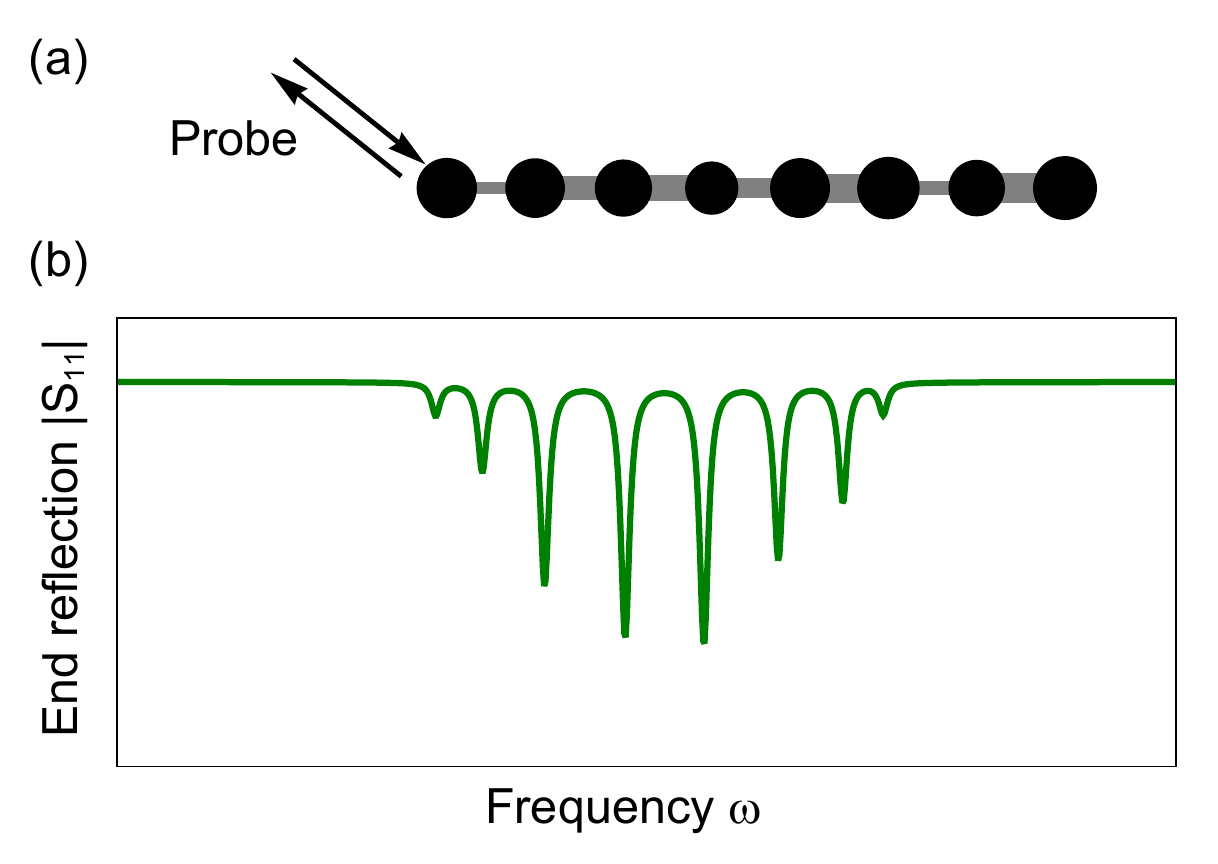}
\caption{\label{Figure:1DTheoryFig} Spectroscopy of a non-interacting 1D chain. (a) A finite-length 1D chain with site-dependent onsite energy (indicated by disk size), and site-dependent tunneling (indicated by connector size). (b) Reflection spectra observed for such a chain.}
\end{figure}

Here we consider a 1D tight-binding lattice, characterized entirely by nearest neighbor tunneling matrix elements $t_\mu$ between sites $\mu$ and $\mu+1$, and onsite energy of site $\mu$, $\delta_\mu$ (see Fig.~\ref{Figure:1DTheoryFig}):
\begin{equation}
H_{1D}=\sum_\mu\left[\delta_\mu a_\mu^\dagger a_\mu-(t_\mu a_{\mu+1}^\dagger a_\mu + t_\mu^*a_\mu^\dagger a_{\mu+1})\right]
\end{equation}

For $n$ lattice sites, this system has $2n-1$ unknowns, coming from the $n$ onsite energies, and $n-1$ tunneling matrix elements; it is thus conceivable that measuring the $n$ eigenmode energies, and $n$ spectral weights (the latter providing $n-1$ linearly independent pieces of information, due to normalization), via a reflection measurement off of a single lattice site, would be enough to extract all system parameters. Symmetry precludes this unless the probed site is at the end of the 1D chain, as proven previously in Burgarth \emph{et al.} \cite{1DTomography2009}.

This prescription allows us to extract all onsite energies $\delta_\mu$ and tunneling matrix elements $t_\mu$, from measured resonance frequencies $\omega^j$ and their spectral weights $\psi_{\mu=0}^j$, normalized such that $\sum^j\left|\psi_{0}^j\right|^2=1$. With measurements only at one end of the chain ($\mu=0$), we obtain all relevant lattice parameters:

\begin{align}
\delta_\mu&=&\sum_{j}{\omega^j \left| \psi_{\mu-1}^j \right |^2}\nonumber\\
|t_\mu|&=&\sqrt{\sum_{j}{\left[(\omega^j-\delta_\mu)\psi_{\mu-1}^j-|t_{\mu-1}|\psi_{\mu-2}^j\right]^2}}\nonumber\\
\psi_{\mu}^j&=&\frac{1}{|t_\mu|}\left[\psi_{\mu-1}^j(\omega^j-\delta_\mu)-|t_{\mu-1}|\psi_{\mu-2}^j\right]
\end{align}

Here we have implicitly assumed $\psi_{\mu=-1}^j=0$ for all $j$. Raised, Roman indices refer to eigenmodes, while lowered, Greek indices refer to sites, counted from the probed end of the chain. Note that the expression for $\delta_{\mu=0}$ reduces to the results from the main text.

\bibliography{LatticeSpec}{}

\end{document}